\documentclass[aps,prc,showpacs,superscriptaddress]{revtex4-1}
\usepackage{times}
\usepackage{amsfonts}
\usepackage{amssymb}
\usepackage{amsmath}
\usepackage{amsopn}
\usepackage{xspace}
\usepackage{bm}
\usepackage{array}
\usepackage{epsfig}
\usepackage{mathrsfs}
\usepackage{endnotes}
\usepackage{natbib}
\usepackage[colorlinks,breaklinks]{hyperref}
\usepackage{nicefrac}
\usepackage{empheq}
\usepackage{graphics}
\usepackage{epsfig}
\usepackage{textpos,colortbl,xcolor}
\usepackage{fancybox,fancyhdr}
\newcommand{\be}{\begin{equation}}
\newcommand{\ee}{\end{equation}}

\newcommand{\ber}{\begin{eqnarray}}
\newcommand{\eer}{\end{eqnarray}}

\newcommand{\bs}[1]{\ensuremath{\boldsymbol{#1}}}
\newcommand{\bea}{\begin{eqnarray}}
\newcommand{\eea}{\end{eqnarray}}
\newcommand{\beq}{\begin{align}}
\newcommand{\eeq}{\end{align}}

\newcommand{\rvec}{\bs{r}}

\newcommand{\bra}{\langle}
\newcommand{\ket}{\rangle}

\begin{document}

\title{The Entanglement Entropy between Short Range Correlations and the Fermi Sea in Nuclear Structure}
\author{Ehoud Pazy}
\affiliation{Massachusetts Institute of Technology, Cambridge, MA, 02139 USA \footnote{On leave from Department of Physics, NRCN, P.O.B. 9001, Beer-Sheva 84190, Israel}
}

\begin{abstract} 
We calculate the nuclear structure orbital entanglement entropy of short range correlations (SRC) based on the nuclear scale separation. Specifically, the entanglement between the SRC orbitals and the rest of the system. It should be stressed that this is a single nucleon not a pair entanglement entropy between the proton and neutron. The entanglement arises from the probability for a nucleon to occupy a momentum state above the Fermi momentum.
We separate the momentum space of the nucleus into two parts  such that nucleons can occupy the meanfield part of the wave function, i.e. Fermi sea (FS) and separately the high-momentum SRC part. The orbital entropy we obtain is between these two parts where we essentially define two momentum subspaces, one containing all the low momentum FS states and the other the high-momentum part as a SRC "orbital" state.
For the calculation we employ the decoupling of low and high-momenta which was established by the similarity normalization group the SRC is viewed as a further "orbital" which can be multiply occupied. 
Since the probability of the occupation of a single SRC is given by the nuclear contact we are able to obtain a simple general expression of the orbital entanglement entropy for SRC by employing the generalized contact formalism. 
This general formula for the SRC orbital entanglement entropy of a nuclear structure in terms of the nuclear contact, allows us to obtain the scaling of the entropy in terms the mass number, $A$. We find that,  unlike the entanglement entropy of many quantum systems which scales with the surface area, the orbital entanglement entropy associated with the SRC in large nuclei is linearly dependent on $A$, i.e., it is shown to be extensive. 
\end{abstract}

\bigskip

%\pacs{25.30.Fj, 21.45.-v, 21.30.-x, 24.30.Cz}

%keywords: 
\maketitle
\section{Introduction}
\label{sec:introduction}
Entanglement entropy, is an entropy arising not from a lack of knowledge, due to thermal fluctuations, of the micro-state the system is in, but is rather an entropy arising due to the entanglement between sub-parts of the system. It arises when one chooses to integrate over part of the system. As such, this entropy can also exists for the ground state. The entropy of the reduced state of a sub-region tends to grow like the boundary area of that sub-region, often with a small, typically logarithmic correction, rather than extensively with the volume \cite{Plenio10}. The scaling of the entanglement with the area of a region is referred to as an area law and holds insights with regard to the distribution of quantum correlations in many-body systems and thus offers a measure of the complexity of the system. Since entanglement is an important resource for quantum computation, sensing, as well as for communication, there is a growing interest in calculating the entanglement entropy of different many-body systems. One obtains area laws for one and higher-dimensional lattices, for fermionic or spin degrees of freedom \cite{Amico08}. The entanglement entropy has been calculated for electronic orbitals \cite{Rissler06} and area laws have been obtained for condensed matter systems \cite{Laflorencie16}. The entanglement entropy is also a good measure for identifying correlations in many-body systems and as such it was also used as a tool for identifying topological order \cite{Jiang12}. 
In addition to the theoretical interest in trying to understand the quantum structure of the nucleus, calculating the entanglement entropy of nuclear structure has practical implications because it can play a role as an organizational principle. Entanglement entropy measures the distribution of wave-function coefficients, and thus it acts as a measure for the viability of truncation of a computational model space. As such it has many implications regarding the reduction of the exponentially large computational basis \cite{Gorton19} where knowledge of the entanglement entropy allows one to require less computational resources while retaining similar accuracy. More specifically, Ref. \cite{Legeza15} demonstrates how employing a quantum entropy based optimization procedure can effectively reduce the number of block states needed to provide an accurate calculation for the ground and first excited state of $^{28}$Si. Recently it has also been shown how nuclear structure can be investigated on a quantum computer where the evaluation of the entanglement entropy plays an important role in assessing the required quantum resources in the number of qubits and circuit depths \cite{Stetcu21}.

The calculation of the entanglement entropy for nuclear structures has commenced with an initial study of the single-orbital entropy and two-orbital mutual information performed for $^{28}$Si, $^{56}$Ni and $^{64}$Ge employing the density matrix renormalization group (DMRG) \cite{Legeza15}. It continued, rather recently, with the calculation for the state of two interacting particles \cite{Kruppa21} and shortly after with a detailed numerical calculation for the entanglement entropy of $^4$He and $^6$He \cite{Robin21}. Mode entanglement has also been investigated under the framework of the Lipkin model \cite{Faba21}. In this work we modify the calculation of the mode (orbital) entanglement by considering the entanglement entropy for a subspace of the system. The subspace for which we calculate the mode entanglement entropy consists of the high-momentum states above the Fermi momentum for the SRC states. It should be stressed that we are not calculating the entropy between the nucleons in the SRC but rather consider the SRC as a further single state which the nucleons can multiply occupy. Due to the universality of the SRC state we are able to obtain a general formula for the entanglement entropy in terms of the high-momentum states rather then having to perform specific calculations for different nuclei. A similar approach of treating the entropy of a Fermi system with a mean-field (MF) part and a correlated part was recently employed in Ref. \cite{Bulgac22}. There the orbital entanglement entropy is given simply by the zero temperature limit of the one-body entropy of a Fermi system. Whereas typically entanglement is used as a diagnostic tool for defining quantum correlations \cite{Klco22}, here this role is reversed and we use the correlations as a basis to evaluate entropy entanglement. This sort of approach allows us also to look at the spatial scaling of the entanglement entropy identifying possible area laws.

Calculating the nuclear structure is highly complex. Due to the repulsive-core of nucleon-nucleon interactions, combined with the large numbers of many-body states, it
was initially considered to be unachievable in a reliable way. It should be noted that it is primarily the entanglement of the many-body nuclear wave function that complicates the calculation since the system can not be well approximated by separable states.
The status of nuclear structure calculations has greatly advanced since the 1970's through the use of modern field theoretic techniques.  Major tools were developed in enabling the study of nuclear structure, e.g. quantum Monte Carlo \cite{Epelbaum11,Carlson15}, Coupled Cluster \cite{Hagen10}, no core Shell Model \cite{Barrett11} and Similarity Renormalization Group (SRG). In the current context we employ the SRG transformation which results in scale seperation. This formalism also helps resolve the following apparent contradiction \cite{Bogner12,Tropiano21}. On the one hand, short range correlations (SRC) have been experimentally observed \cite{Egiyan03,Egiyan06,Subedi08,Fomin12,Hen17,Torres20,Torres21,Piasetzky06} in which one observes components in the high resolution nuclear wave function with relative pair momenta greater than the Fermi momentum \cite{Korover14,HenSci14,Korover21,Duer18,Duer19,Schmookler19,Schmidt21}. On the other hand, the shell model for the nucleus, which has been highly successful for calculating nuclear properties \cite{deShalit63} seems to include no such short range structure thus seemingly contradicting the SRC description. Through SRG it was recently demonstrated \cite{Bogner12,Tropiano21} that these two viable seemingly contradicting descriptions for the nucleus, the FS like independent particle model and SRC can in fact be connected. This sort of connection has also been previously studied and demonstrated for the $^4$He case \cite{Neff15}. Employing a SRG transformation specifically enables one to obtain a factorization of the long from short-distance physics, providing a clean scale separation. The SRC are manifested when considering high RG resolution. Whereas in the low-RG resolution the features of SRC phenomenology are clearly identified using simple two-body operators and local-density approximations with uncorrelated wave functions reconciling the contrasting pictures. One can thus naively view SRC as the two-body extension to the single independent particle form of the shell model. In this naive view the physics of the nucleus is encapsulated in one plus correlated two-body model which are scale separated. This separation of scales is extremely effective for the calculation of the entanglement entropy of the nucleus. In the present work we exemplify how the mode entanglement entropy is given in terms of SRC instead of considering single particle states above the Fermi surface we consider the single particles to be occupying a universal SRC state. For simplicity, we only consider the SRC composed of neutron-proton pairs which are in a $s$-wave configuration. The extension to other isospin SRC is more elborate, however, neutron-proton pairs have been found experimentally to constitute the vast majority of SRC. In only considering such pairs we in fact neglect the contribution of neutron-neutron and proton-proton pairs to the entanglement entropy. However this contribution has been shown to be small. In fact in the electron scattering experiments kinematically designed to measure SRC a typical ratio of proton-neutron SRC pairs to proton-proton SRC pairs was found to be around 20 \cite{Subedi08,Hen17,Piasetzky06,Korover14,HenSci14,Korover21,Duer18,Duer19}. 

We start by a brief reminder of the orbital entanglement entropy given in Sec. \ref{sec:orbitalent}. In Sec. \ref{sec:ent_SRC} the generalized contact formalism (GCF) is succinctly introduced.
We then extend the definition of the orbital entanglement entropy to include an entanglement entropy for a subspace of the momentum space which we define as the "orbital" entanglement entropy of the SRC. The orbital enatnglement entropy of the SRC is then analytically calculated in terms of the nuclear contact. In Sec. \ref{sec:comparison} we compare our results with previously obtained results from the literature for the entanglement entropy of $^4$He \cite{Robin21} and those for the (one-body) entropy of a Fermi system \cite{Bulgac22} . The results for the entanglement entropy for the SRC of the nuclear structure are summarized and future research directions are discussed in Sec. \ref{sec:conclusion}. 

\section{Orbital entanglement entropy}
\label{sec:orbitalent}
The entanglement entropy measures the correlation between two parts of a bipartite system \cite{Plenio07,Horodecki09}. In the case of distinguishable particles, the notion of the
entanglement is based on the structure of the tensor product for the Hilbert spaces of the subsystems. However, trying to define a similar concept for identical particles is not straightforward, since the decomposition into particle subsystems does not correspond to a tensor product structure of the Fock space. In order to address this issue, the notion of the mode (orbital) entanglement was introduced \cite{Zanardi02,Gittings02,Shi03}. 

One can define the eigenstates of the nucleus, $|\Psi \ket$, by choosing a single particle basis defined by the quantum numbers $\{i\}=\{n_i,J_i,l_i,m_i,s_i,\tau_i\}$, which correspond to the principal quantum number, the total angular momentum, the orbital angular momentum, and the spin, isospin projections. The nucleus eigenstates can then be written as a linear combination of Slater determinants $|\phi \ket$ for the nucleon wave functions 
\be
\label{eq:nuclearwf}
|\Psi \ket=\sum_\eta  \mathcal{A}_\eta |\phi_\eta \ket,
\ee
where the Slater determinant is given in terms of applying creation operators on the true particle vacuum $|0\ket$
\be 
\label{eq:slater}
|\phi_\eta \ket =\prod_{i \in \eta}^{A} a_i^\dagger|0 \ket,
\ee
where $A$ is the number of nucleons in the nucleus. 

Employing this notation, the single orbital entanglement entropy $S_i^{(1)}$ is given in terms of the one-orbital density matrix \cite{Robin21} 
\be
\label{eq:one_density_mat}
\rho^{(i)}=
\begin{pmatrix} 
1-\gamma_{ii} & 0 \\
0 & \gamma_{ii}
\end{pmatrix},
\ee
where the occupation of orbital, $i$, is given by $\gamma_{ii} =\bra \Psi| a_i^{\dagger} a_i|\Psi \ket$. 

To obtain the entanglement entropy essentially one needs to perform a calculation of the von Neumann entropy of the partial density matrix, Eq. (\ref{eq:one_density_mat}) 
\begin{equation}
\label{eq:one_orb_ent}
S_{i}^{(1)}=-Tr[\rho^{(i)}\ln \rho^{(i)}]=-\sum_{k=1}^2 \omega_k^{(i)} \ln \omega_k^{(i)},
\end{equation}
where $\omega_k$ are the eigenvalues of $\rho^{(i)}$

For the sake of completeness we also present the two-orbital entanglement entropy which is similarly given by 
\begin{equation}
\label{eq:two-orbital}
S_{ij}^{(2)}=-Tr[\rho^{(ij)}\ln \rho^{(ij)}]=-\sum_{k=1}^4 \eta_k^{(ij)} \ln \eta_k^{(ij)},
\end{equation}
where $\rho^{(ij)}$ is the following $4 \times 4$ matrix with eigenvalues $\eta_k^{(ij)}$ \cite{Robin21}
\begin{equation}
\label{eq:twoptdenmat}
\rho^{(ij)}=
\begin{pmatrix}
&1-\gamma_{ii} -\gamma_{ij}+\gamma_{ijij} & 0 & 0 & 0 \nonumber \\ 
& 0 & \gamma_{jj}-\gamma_{ijij} &  \gamma_{ij} & 0  \nonumber \\
& 0 & \gamma_{ij} & \gamma_{ii}-\gamma_{ijij} & 0  \nonumber \\
& 0 & 0 & 0 & \gamma_{ijij}
\end{pmatrix},
\end{equation}
defined in terms of  $\gamma_{ij} =\bra \Psi| a_j^{\dagger} a_i|\Psi \ket$ and 
$\gamma_{ijij}=\bra \Psi| a_i^{\dagger}a_j^{\dagger}a_j a_i|\Psi \ket$.

\section{Entanglement entropy in terms of the nuclear contact}
\label{sec:ent_SRC}
Having introduced the concept of orbital entanglement entropy, we modify it by considering the entanglement entropy for the subspcae of high-momentum states. We consider this subspace as aseparate "orbital" , SRC state, which can be multiply occupied. In this format the mode entropy for high-momentum states is given in terms of SRC instead of single particle states. We show specifically that the splitting of the wave function into two parts, one part is a MF part and the other part is an SRC part which is defined by a single two-body wave function in space with undefined center of mass cordinate. This separation
which is the basis of GCF, provides a very convenient framework for this calculation. One does not need to know the specific wave function for the SRC pair just to realize it is universal and as such it is its occupation that defines its entanglement entropy. 

The GCF fundamentally complements the shell model enabling one to incorporate SRC into the model. Though the shell model is essentially a MF theory, it provides detailed information about the nuclear shell structure. One of the major success of the shell model is  explaining the stability of a certain number of nucleons in nuclei called the magic numbers. However, it does not capture the complete picture of the nucleus structure. The shell model fails to predict the occupancy of the shells since in particular it does not describe the strong short range nucleon-nucleon part of the potential which is responsible for the SRC. The SRC are manifested in a high-momentum tail of the nucleon momentum distribution with momenta exceeding the Fermi momentum. Experimental evidence for the existence of SRC was obtained by performing electron scattering experiments. SRC related physics was observed by choosing the right kinematics in these experiments, specifically inclusive scattering at Bjorken $x_B>1$ \cite{Egiyan03,Egiyan06,Fomin12} and also by performing exclusive experiments in which the two-body currents could be identified and distinguished from final state interactions \cite{Piasetzky06,Korover14,HenSci14}.

Starting off with the ground state of the nucleus with hard interactions, i.e., low-RG resolution, there are both high-momentum and low-momentum contributions to the wave function. By employing a SRG transformation to high-RG resolution it was demonstrated that operator expectation values exhibit factorization in the two-nucleon system \cite{Bogner12,Tropiano21}. Through this factorization a model can be obtained in which SRC, which are identified as components of the nuclear wave function in which nucleon pairs with relative momenta well above the Fermi momentum, naturally emerge. This factorization is the basis of the GCF \cite{WeissPRC,Weiss15,Weiss17,Weiss16}, which is a model introduced to provide a simple framework for describing the SRC. The GCF is a very convenient theoretical tool for analyzing experiments designed to probe SRC in the nucleus \cite{HenSci14,Alvioli13,ArrHigRos12,Wiringa14}. It has been benchmarked against ab-initio many-body calculations, proving its validity, and was successfully applied to a wide range of topics, most notably to the analysis of electron scattering measurements \cite{Weiss18,Weiss19,Duer19,Pybus20,Weiss21}. The contact which is at the heart of the GCF will be shown to be an instrumental tool for expressing the SRC entanglement entropy.
Currently the GCF does not treat three-body effects. However, the three-body contributions are expected to be less important than those of the leading two-body.

\subsection{The nuclear contact}
\label{subsec:contact}
The GCF is based on the factorization ansatz of the many-body wave function into a two-body problem of a nucleon pair close in space (correlated), which is a universal part, common to all nuclei, multiplied by a particular part, that depends on the specific state. The specific state part
describes the remaining, $A-2$, nucleons in the nucleus \cite{WeissPRC} and also depends on the center of mass coordinate for the nucleons pair
\be
\label{eq:ansatz}
  \Psi\xrightarrow[r_{ab}\rightarrow 0]{}
             \sum_\alpha\varphi_{ab}^\alpha(\rvec_{ab})
             A_{ab}^\alpha\left(\bs{R}_{ab},\{\rvec_c\}_{c\not=a,b}\right).
\ee
Here $\varphi_{ab}^\alpha$ are the two-body universal functions defining the SRC, $A_{ab}^\alpha$ are the so called regular part of the many-body wave function, the index $\alpha$ defines the quantum numbers for the two-body states and the indices $ab$ define the SRC in terms of isospin, $pp$, $pn$ and $nn$ pairs as well as the specific particle indices in $\rvec_{ab}$ and 
$\bs{R}_{ab}$. This sort of approximation is viable when the wave function is viewed on a short length scale, important for calculating short range observables. For simplicity and since most experimental data on SRC is in the kinematical region where neutron-proton pairs dominate \cite{Subedi08,Hen17,Piasetzky06,Korover14,HenSci14,Korover21,Duer18,Duer19}, we consider in this work only neutron-proton pairs in the deuteron like quantum state (quasi-deuteron). These pairs are defined by $\alpha \equiv (S=1, \pi=+1, J=1, M=0 \pm 1)$, where $S$ is the total spin of the pair, $\pi$ is their parity, and  $J$ and $M$ are the quantum numbers defining the SRC, which are the total angular momentum of the pair and its projection respectively. This state has been shown to match experimental data to a high precision \cite{Subedi08,Hen17,Piasetzky06,Korover14,HenSci14,Korover21,Duer18,Duer19}. Since we will be considering only this proton neutron quantum state for the SRC pair we will for simplicity eliminate $pn$ index aswell as the quantum state index $\alpha$.
Under this approximation the nuclear contact in GCF is defined as
\begin{equation} 
\label{eq:contact}
C= N(A,Z)\bra A | A\ket,
\end{equation}
where $N(A,Z)$ is the number of pairs one can produce from $Z$ protons and, $A-Z$, neutrons. In this work we will only consider symmetric nuclei such that $Z=A/2$.

The function $\varphi_{ab}(\rvec_{ab})$ in Eq. (\ref{eq:ansatz}) is a function of solely the distance between the SRC proton and neutron, $\rvec_{ab}$, and not their center of mass coordinate $\bs{R}_{ab}$ which appears in $A_{ab}\left(\bs{R}_{ab},\{\rvec_c\}_{c\not=a,b}\right)$. This function is obtained by solving the two-body, zero energy, Schr\"{o}dinger equation with the full many-body potential. 

Switching to momentum states the one nucleon momentum distribution
\be
\label{eq:occupency}
n({\bf q})= \bra \Psi | a_{\bf q}^\dagger a_{\bf q}| \Psi\ket .
\ee
can be expressed by employing the contact Eq. (\ref{eq:contact}) for large momentum limit  $|{\bf q}|\ll k_F$. It is well approximated by \cite{WeissPRC}
\be 
\label{eq:momentumoc}
n({\bf q})=C|\phi({\bf q})|^2,
\ee
where $\phi({\bf q})$ is the Fourier transform of $\varphi_{ab}(\rvec_{ab})$ in Eq. (\ref{eq:ansatz}) and the normalization for, $n({\bf q})$ for a $N=Z$ nucleus 
is given such that, $n({\bf q})$, gives the fraction of the one body momentum above $k_F$ \cite{Weiss18}  
\be 
\label{eq:Cprob}
{\int_{k_F}^\infty n({\bf q}) d{\bf q} \over\int_0^\infty n({\bf q}) d{\bf q}}={C \over(A/2)}
\ee

Since the contact is obtained by tracing out the degrees of freedom exterior to the SRC pair, it is clear why it plays an important part in determining the entanglement entropy of a SRC pair with respect to the rest of the nucleus structure.

\subsection{Orbital entanglement entropy of SRC in terms of the contact}
\label{subsec:SRC entropy}
To calculate the entanglement entropy between the SRC and the rest of the system we define the momentum states with $q>k_F$ as an SRC state which is a relatively good approximation even though excitations near the Fermi level are due to long range correlations \cite{Weiss19}. Thus we divide the momentum space into two distinct regions, the high-momentum region defining the SRC and the mean field region defined by, $q<k_F$. We will further treat the high-momentum states as a single  SRC "orbital" which can be multiply occupied, since we have defined it by all high-momentum states. 

In calculating the entanglement entropy of a particle occupying a momentum state above the Fermi energy, we essentially extend the notion of orbital entanglement entropy to that of a subspace entanglement entropy. We divided the momentum space into two separate sub spaces, $\Phi^i$ which includes all single particle states in which $q<k_F$, i.e. FS, and $\Phi^i_\perp$ which includes all single particle states in which $q>k_F$ which is essentially what we refer to as the SRC "orbital". Defining a projection operator $\hat{P}$ which projects onto states in $\Phi^i_\perp$ we define the analog of $\gamma_{ii}$ in Eq. (\ref{eq:one_density_mat}) to be $\gamma_{_{SRC}}=\bra \Psi |\hat{P}| \Psi \ket$.

To calculate the SRC orbital entanglement entropy one first has to calculate the occupancy probability for a single SRC $\gamma_{_{SRC}}$ which is defined through Eq. (\ref{eq:Cprob})

\be
\label{eq:normalized contact1}
\gamma_{_{SRC}}= {C\over(A/2)}\equiv c.
\ee
Where we have defined for convenience a normalized contact $c$ which to a good approximation does not depend on $A$.
 
Employing the SRC occupancy, Eq. ({\ref{eq:normalized contact1}), a one "orbital" SRC density matrix, $\rho^{(SRC)}$ as in Eq. (\ref{eq:one_density_mat}) can be obtained
with the following eigenvalues: $\omega_k^{(SRC)}=\{1-c ,c\}$. 
The orbital entanglement entropy for a single SRC is directly obtained through Eq. (\ref{eq:one_orb_ent}) 
\be 
\label{eq:Ent_SRC}
S^{SRC}=- \left [c \ln\left ({c \over 1-c }\right )+\ln{(1-c)} \right ].
\ee
It should be noted that under the normalization, Eq. (\ref{eq:Cprob}) the probability of a particle being in FS is given by
 $\gamma_{_{FS}}=1-\gamma_{_{SRC}}$. Thus, as should be expected, the eigenvalues for the FS are given by $\omega_k^{(FS)}=\{c ,1-c\}$. So that the orbital entanglement of the whole FS subspace is also given by Eq. (\ref{eq:Ent_SRC}).

To obtain the total SRC orbital entanglement entropy one has to multiply the single SRC entanglement, Eq. (\ref{eq:Ent_SRC}) entropy by plugging the expression for the number of nucleons   a certain nucleon can be paired with $N(A,Z)=A/2$ as expressed in Eq. (\ref{eq:contact}). Thus the total SRC entanglement entropy is obtained,
\be 
\label{eq:Ent_tot_SRC}
S_{A}^{SRC}= -{A\over 2} \left [c  \ln\left ({c \over 1-c }\right )+\ln{(1-c) }\right ].
\ee
This simple general analytic expression for the orbital entanglement of the SRC part of the orbital entanglement entropy in terms of the reduced contact is viable to any nucleus under the approximation that SRC are  defined by the subspace of states with momentum above the Fermi momentum.

\subsection{Scaling of the SRC entanglement entropy}
\label{subsec:scaling}
The above result, Eq. (\ref{eq:Ent_tot_SRC}),  shows an explicit linear dependence of the SRC entanglement entropy on $A$. This is a rather surprising result since the
SRC entanglement primarily scales like the nuclei volume $A$, making the entanglement entropy extensive which means that it does not obey an area law as could be expected. This extensive scaling of the entanglement entropy can be traced back to the number of available nucleons a nucleon can pair with  to form a SRC, $N(A,Z)\propto A$. There might be a further dependence on $A$ through the normalized contact, $c$. However, the normalized contact is expected to be $A$ independent for nuclei with a large $A$ for which the Fermi gas approximation is expected to be valid. The linear dependence of the contact on $A$ and the independence of the reduced contact in the Fermi gas approximation can be inferred from Ref. \cite{Weiss19b}. There the contact was calculated by taking the limit when interparticle distance $r$ goes to zero of the two-body Fermi gas correlation function $g(r)$. The obtained result for the Fermi gas model was shown to be linearly dependent on the number of nucleons in the nucleus and the slope of the linear dependence on $A$ was then estimated by comparison to experimental results as well as to calculations based on the AV18 potential \cite{Weiss19b}.
The obtained result for the reduced contact for a symmetric nuclei was given as $c_{pn}^{FG}=0.146 \pm 0.002$. Thus under the Fermi gas approximation the SRC entanglement entropy, Eq. (\ref{eq:Ent_tot_SRC}), is simply linear in $A$.

The reduced contact was also calculated using a variational Monte Carlo (VMC) method and calculations were compared to experimental results and found to be in agreement.
Through the VMC results one can also justify numerically the neglect of other channels differing form the proton-neutron channel $(S=1, \pi=+1, J=1, M=0 \pm 1)$. The proton-proton and neutron-neutron calculated contacts were shown to be an order of magnitude smaller then the proton-neutron channel and the $S=0$ proton-neutron channel is even smaller. The VMC calculated values in momentum space, for  the reduced contact in Ref. \cite{Weiss18}, range from $c_{pn} \approx 0.125$ for $^4$He, $^8$Be, $^9$Be, $^{10}$B to $c_{pn} = 0.106$ for $^6$Li, $^7$Li and a higher value $c_{pn}= 0.168$ for $^{12}$C. The experimental values obtained for $^4$He were slightly higher than the VMC calculation $c_{pn}^{exp} \approx 0.149$ as was the experimental evaluated contact for $^{12}$C which was $c_{pn}^{exp} \approx 0.168$. It is difficult to determine if there is a slight mass number dependence of the reduced contact due to the limited range of nuclei calculated. As a rough approximation one can take the value for the contact as $c_{pn} \approx 0.15$ and as independent of $A$. However, it seems that there could be some corrections to the simple linear (extensive) dependence of the entanglement entropy on $A$.

A word of caution should be given when estimating the value for the reduced contact by calibrating the result through experimental data. Whereas the relevant scale for momentum separation is well defined theoretically by SRG it is not always clear that this was the relevant scale in the experiments. The obtained experimental results depends on physical scale through the specific kinematics. These two scales, the theoretical and experimental, need not necessarily coincide. This issue is elaborated in detail in \cite{Tropiano21}. The scale mismatch if occurring can cause errors in estimating the reduced contact and thus also effect the entanglement entropy estimation through Eq. (\ref{eq:Ent_tot_SRC}).

\section{Comparison to previous results}
\label{sec:comparison}

Calculating the entanglement entropy for a nuclei can involve considerable computational efforts when considering single particle energy levels. As such it can only be preformed for few-body systems, due to the extensive computational effort which scales rapidly due to combinatorics. In the following section we will compare our results to such a few-body calculation \cite{Robin21}. It should however be noted that we have calculated the "orbital" entanglement entropy of a subspace of high-momentum states rather then a single orbital, moreover one should also note that 
 the orbital entropy calculation is basis dependent and as such the comparison  to the few-body results should be taken with a grain of salt. In addition below we present a comparison with respect to a calculation of the one body entropy which was calculated for a Fermi system with a MF part and a correlated part \cite{Bulgac22}, this calculation used the same basis we have however in a different method. We demonstrate how taking the zero temperature limit for the one body entropy one and summing over the high-momentum states one obtains the same result.

The sole source for an entanglement entropy calculation involving single particle energy levels which we found easy to compare to quantitatively was the calculation performed for $^4$He and $^6$He \cite{Robin21}. The calculation of the entanglement entropy was performed by a full diagonalization on a 7 shell model space. It should be noted that the calculation that was performed with only 6 major active shells, containing the first 114 single-particle states, was not large enough to give satisfactory results. On the other side it seems rather simple using our SRC entanglement expression Eq. (\ref {eq:Ent_tot_SRC}) to calculate the SRC entanglement entropy once the relevant reduced contact is known. Though one should keep in mind that we have only considered a subspace SRC "orbital" considering the orbital entanglement entropy resulting from the distinction between FS or the SRC states. It is thus of interest to try to evaluate how much of the entanglement entropy can be related to this seperation. In this section we compare the results for the entanglement entropy of $^4$He from Ref. \cite{Robin21} to the SRC entanglement entropy from our analytical expression with the appropriate reduced contact and mass number for $^4$He.

In Ref. \cite{Robin21} a comparison was made between different basis states to find the appropriate base in which to work in. It was found that the two most accurate basis sets were NAT, a term for natural, and VNAT, a term for, for variational natural  which gave the most accurate results. The total sum of the single orbital entanglement entropy for the calculations in these basis gave a result of $S_{tot}^{(1)}(^4He) =1.006$. In calculating the single-orbital entanglement entropy for $^4$He solely from Eq. (\ref{eq:Ent_tot_SRC}) we obtain a value of $S^{SRC}(^4He)=0.74$ for $c_{pn}=0.12$ and $S^{SRC}(^4He)=0.86$ for the value for the reduced contact obtained experimentally $c_{pn}^{exp}=0.15$ \cite{Weiss18}. Surprisingly, the results are relatively similar to those of  Ref. \cite{Robin21} especially keeping in mind the simple SRC "orbital" like formalism. Moreover, contributions from proton-proton and neutron-neutron SRC pairs were not considered as well as two-orbital entanglement entropy involving pairs of nucleons. Taking this comparison literally one can claim that probably much of the nuclear entanglement entropy is due to the SRC separation from the FS, i.e., that the main source for the entanglement entropy is obtained by the possibility for nucleons to excited to high-momentum states.
However, due to the use of a different basis for the calculations, one can only view the relative agreement between them as a trend. One should keep in mind that the calculation of the entropy entanglement in \cite{Robin21} was performed for the low-RG case, whereas the results obtained in this work, Eq. (\ref{eq:Ent_tot_SRC}) are for the high-RG case. In Ref. \cite{Tropiano21} a SRG connection was established showing that calculations performed in the high-RG resolution involving SRC pairs in the GCF can be mapped via SRG to the low-RG resolution were he nucleons are best described by a FS. It still needs to be shown how the SRG connects these two methods for calculating the entanglement entropy and how it continuously transform from one to the other. An indication for this connection can be found in the claim made in Ref. \cite{Robin21} that the most important couplings are $1s$-$1s$, $1s$-$1p_{1/2}$ and $1p_{1/2}$-$1p_{1/2}$ since they are related to deuteron-type correlations. A further qualitative comparison can also be made to Ref. \cite{Legeza15} in which the single orbital and two-orbital mutual information was calculated employing the DMRG for  $^{28}$Si, $^{56}$Ni and $^{64}$Ge. The total entanglement entropy for these calculations was not presented thus allowing only a qualitative comparison. The authors of of Ref. \cite{Legeza15} state that they could see significant entanglement between proton-neutron maximally aligned states for the $p$3/2 and $f$ 5/2 orbits. However, it is also claimed that the fact that the two orbital mutual information is approximately equal for the $p$1/2, $p$3/2, and $f$ 5/2 orbits,
and independent of their $M$ projections, is an indication for the presence of a strong $T=1$ proton-proton and neutron-neutron pairing coherence. Whereas in our calculations we have neglected such SRC pairs. Future work should establish how the entanglement entropy is effected by the SRG transformation.

A calculation  for the nuclear orbital entanglement entropy involving SRC was recently preformed in Ref. \cite{Bulgac22}. The starting point for that calculation was the one-body Fermi system Boltzmann entropy defined in terms of the fermion momentum distribution $n(k)$ 
\be
\label{eq:onebodyEnt}
S=-g \int {dk \over {(2 \pi)}^3} n(k) \ln n(k)-g \int {dk \over {(2 \pi)}^3} [1-n(k)] \ln [1-n(k)],
\ee
where in the nuclear case $g=4$ is the spin-isospin degeneracy factor. The one-body entropy was calculated depending on the temperature from which one can obtain the orbital entanglement entropy by taking the zero temperature limit. The SRC were incorporated into the calculation by considering the following nucleon momentum distribution
\be
\label{eq:nucelondis}
n(k) = \eta(k_0)
                          \begin{cases}		
				n_{mf}(k)  & \text{if}  \  k \le k_0 \\
		 		{C_{pn} \over k^4}    & \text{if} \ k_0  <k < \Lambda,
			\end{cases}
\ee
where $\Lambda$ is a cut-off and will be assumed $\Lambda=\infty$.
The result for the SRC orbital entanglement entropy can be obtained from Eqs. (\ref{eq:onebodyEnt}, \ref{eq:nucelondis}) by integrating over all momentum states above the Fermi momentum. This result is matched to the SRC entanglement entropy as defined by Eq. (\ref{eq:Ent_tot_SRC}) the details are described in the  Appendix.

\section{Summary and Discussion}
\label{sec:conclusion}
Basing our work on previously obtained theoretical results demonstrating that through applying an SRG transformation on the nuclear structure one obtains scale separation \cite{Tropiano21}, we have calculated the nuclear structure orbital entanglement entropy for SRC. It should be stressed that the SRC orbital entanglement entropy we have calculated is a single nucleon entropy related to the probability of finding a nucleon in the FS or as part of an SRC pair. This is a rather simplified model in which the SRC is identified with the high-momentum subspace and considered as a single "orbital". Thus a nucleon can occupy either one of the FS orbitals or an SRC pair. Essentially, based on the scale separation we decoupled the Hilbert space such that
\be
\label{eq:Hilbert_space}
\mathcal{H}=\mathcal{H}_{FS}\otimes \mathcal{H}_{SRC}.
\ee
It is a product of a FS Hilbert subspace, $\mathcal{H}_{FS}$ and the SRC Hilbert subspace $\mathcal{H}_{SRC}$. Based on this decomposition, Eq. (\ref{eq:Hilbert_space}) we calculated the reduced density matrix for the SRC momentum subspace
\be
\label{eq:reduced-density_matrix}
\rho_{_{SRC}}=\text{Tr}_{_{FS}} \rho
\ee
where $\rho=|\Psi \ket \bra \Psi|$. 
Employing  the GCF formalism allowed us to obtain a simple analytic expression for the reduced density matrix, Eq. (\ref{eq:reduced-density_matrix}), in terms of the reduced nuclear contact, Eq. (\ref{eq:normalized contact1}) and through it to obtain a simple analytical expression for the entanglement entropy, Eq. (\ref{eq:Ent_tot_SRC}). Through this general formula we have been able to demonstrate that the orbital entanglement entropy for the high-momentum states, i.e., the SRC orbital part of the nuclear structure, scales extensively and not according to an area law. 

More specifically, we utilized in our calculations the fact that in the high-RG resolution the nuclear structure can be viewed as a bipartite system where the high-momentum short length scale properties of the system are given in terms of SRC pairs. Based on this we have performed our calculation for the SRC entanglement entropy in the framework of the orbital entanglement entropy, in which the entanglement is given by the probability an "orbital" is occupied. In the calculation we considered the high-momentum states as the SRC part of the many-body wave function, which can be occupied by many pairs. The reduced nuclear contact for a SRC pair which is obtained by integrating out the long length scales in the nucleus corresponding to the low-momentum degrees of freedom and then normalizing the result, naturally appears as the probability measure for finding  a single SRC pair. Using this connection, one can define a single orbital SRC density matrix , Eq. (\ref{eq:reduced-density_matrix}) in terms of the reduced nuclear contact, Eq. (\ref{eq:normalized contact1}). From the SRC density matrix one can obtain the SRC entanglement entropy through the von Neumann entropy, Eq. (\ref{eq:Ent_SRC}). Summing over the entanglement entropy for all SRC pairs we obtained that the nuclear orbital entanglement entropy for the high-momentum, i.e.,  SRC part, Eq. (\ref{eq:Ent_tot_SRC}). One can notice that it obeys an extensive entropy law, since it scales linearly with the mass number $A$ and since the nuclear structure density is essentially constant this means it scales like the volume of the nucleus. 

Even though in obtaining our results quite a few contributions were neglected, e.g., neutron-neutron and proton-proton SRC still we find them to be relatively close to the more elaborate work performed in calculating the entanglement entropy for $^4$He \cite{Robin21}. This can be considered as a coincidental agreement between the results since in considering the SRC as a single orbital we summed over all high-momentum states. However,  it might be seen as an indication that most of the nuclear entanglement entropy is related to the ability to excite nucleons to high momentum states as part of SRC and thus our general Eq. (\ref{eq:Ent_tot_SRC}) is a good measure for the entanglement entropy of all nuclei.

The work presented here is essentially preliminary. It introduced the idea that the SRG scale separation is an invaluable tool for performing the calculations for the nuclear structure orbital entanglement entropy and demonstrating the scaling of the SRC orbital entanglement entropy. Further research has yet to incorporate proton-proton and neutron-neutron SRC and establish how the orbital entanglement entropy is modified by the SRG transformation.

\section*{Acknowledgments}
\label{Sec:ack}

I wish to thank C. W. Johnson and O. Gorton for sharing their results, C. Robin for discussions. R. Weiss ,G. A. Miller and A. Bulgac  for their valuable advice and comments. Special thanks to O. Hen and N. Barnea for insightful discussions, invaluable comments and Or for his generous hospitality at the LNS which made this work possible. This work was supported by the PAZY Foundation.

\section*{Appendix}
\label{sec:App}
A calculation of the one-body entropy was preformed in Ref. \cite{Bulgac22} for which the zero temperature case corresponds to the orbital entanglement entropy. The SRC were incorporated into that calculation by considering the nucleon momentum distribution described in Eq. (\ref{eq:nucelondis}) choosing as we did here that the distinction in momentum between SRC and the bulk FS is given by the Fermi momentum $k_0=k_F$. For this choice one obtains $\eta(k_F)=0.25$. The contact in this case is given by \cite{Bulgac22}
\be 
\label{eq:contactonebody}
C(k_F)=\eta(k_F) n_{mf}(k_F) k_F^4.
\ee
The SRC entanglement entropy is obtained by considering all momentum states above the Fermi momentum as the SRC "orbital". Thus
Inserting the expression for $n(k>k_F)$ from Eq. (\ref{eq:nucelondis}) into the equation for the entanglement entropy, Eq. (\ref{eq:onebodyEnt}) and approximating the integrals one obtains 
\be
\label{eq:app}
S \approx -{g  \over 2 \pi^2} \left [ {C\over {k_F}} \ln \left ({C \over {k_F}^4 } \over {1-{C \over {k_F}^4}} \right)+{k_F}^3 \ln \left ( 1- {C\over {k_F}^4} \right ) \right].
\ee
Associating  the reduced contact with $c_{pn}={C/ {k_F}^4}$ one obtains 
\be
\label{eq:entapprox}
S \approx -3 A \left [ c_{pn} \ln \left (c_{pn} \over {1-c_{pn}} \right)+ \ln \left ( 1- {c_{pn}} \right ) \right]
\ee
where the different prefactor with regards to Eq. (\ref{eq:Ent_tot_SRC}) is due to the difference in defining the fraction of  which in Eq. (\ref{eq:Cprob}) was simply given by $A/2$ whereas in \cite{Bulgac22} it is given by 
\be
\label{eq:probk}
{n(k>k_F)\over n_0}={3 C(k_F) \over {k_F}^4}.
\ee
 This difference can be related to the number of SRC channels, e.g., proton-proton, neutron-proton, considered.

%=============================================================================

%=============================================================================
\end{document}